\documentclass[english]{lni}

\IfFileExists{latin1.sty}{\usepackage{latin1}}{\usepackage{isolatin1}}

\usepackage[utf8]{inputenc}
\usepackage{latexsym}
\usepackage{ngerman}
\usepackage{graphicx}
\usepackage{array}
\usepackage{multirow}
\usepackage{hhline}		
\usepackage{subfigure}
\usepackage{mathptmx}
\usepackage{mathtools}
\usepackage{amsmath}
\usepackage{amsfonts}
\usepackage{amssymb}
\usepackage{amsbsy}
\usepackage{amsthm}
\usepackage{todonotes}
\usepackage{listings}
\usepackage{color} 
\usepackage{float}
\usepackage[justification=centering]{caption}
\usepackage[export]{adjustbox}
\usepackage{cite}
\usepackage[flushmargin,hang]{footmisc}
\usepackage{tabularx}
\newcolumntype{C}[1]{>{\centering\arraybackslash}p{#1}} 

\author{
	Peter Hillmann, Frank Tietze, and Gabi Dreo Rodosek \\
	\\
	Chair for Communication Systems and Network Security\\ 
	Faculty of Computer Science\\ 
	Universität der Bundeswehr München\\ 
	Werner-Heisenberg-Weg 39\\
	D-85577 Neubiberg\\ 
	\{peter.hillmann, frank.tietze, gabi.dreo\}@unibw.de\\
}
\title{Strategies for Tracking Individual IP Packets\\ Towards DDoS}
\begin{document}
\maketitle

\begin{abstract}
The identification of the exact path that packets are routed in the network is quite a challenge. This paper presents a novel, efficient traceback strategy in combination with a defence system against distributed denial of service (DDoS) attacks named \textit{Tracemax}. A single packets can be directly traced over many more hops than the current existing techniques allow. It let good connections pass while bad ones get thwarted. Initiated by the victim the routers in the network cooperate in tracing and become automatically self-organised and self-managed. The novel concept support analyses of packet flows and transmission paths in a network infrastructure. It can effectively reduce the effect of common bandwidth and resource consumption attacks and foster in addition early warning and prevention.
\end{abstract}

\section{INTRODUCTION}\label{introduction}
One of the most publicly recognized DDoS attack took place in 2007 against Estonian organizations. In such a modern country with an extended e-government, this attack had a large influence on the general public \cite{estonia2007}. In 2010 as protest reaction on inhibition of WikiLeaks bank accounts, MasterCard, Visa and PayPal as well as other large finance institutes got attacked using DDoS attacks \cite{frobes10}. The coordinated action called \glqq Operation Payback\grqq{} has stopped aforesaid services for several hours causing great financial loss.

The power and danger of distributed denial of service (DDoS) attacks are evident and become an increasingly frequent problem of the global Internet. Strong defence mechanisms are necessary since the attacks are focussed not only to single services but entire countries and infrastructures. In 2014 the largest detected DDoS attack had consumed a bandwidth of over 500 GBits/s\cite{Forbes2014}. Nowadays, no service infrastructure with corresponding network nodes is able to process such an amount of data. The attack traffic may originates from a large number of sources like botnets, protest groups or military forces. DDoS attacks are very hard to counter because they do not  target a specific vulnerability of systems and the initiators remain anonymous. 

It is mandatory for many defence strategies to identify the multiple sources of such an attack. In order to detect the sources and to reveal spoofed IP addresses a sophisticated traceback system is needed. The trustworthiness and precision of a traceback technique implemented in the aforesaid system is of highest importance. This enables the identification of the attacker for law enforcement as well as forensic analysis. The presented traceback technique can also be used for identification of hidden channels and routing anomalies as well as input information for intrusion detection systems (IDS). Load balancing systems can be verified and zone routing can be proofed as well.

\section{Problem Description}\label{scenario}
The necessity of an efficient traceback strategy with minimal overhead in transit autonomous systems (Transit AS) and stub autonomous systems (Stub AS) of ISPs is illustrated by using the following typical real world scenario. A DDoS attack and malicious traffic can very easily be detected at the victim \cite{Li2003,DFN2015} using IDS's with specific attack signatures, for example inbound traffic or network consumption delay thresholds. Thereby, fake source IP addresses have to be anticipated. The IDS then provides a true positive alarm and detection vector triggers the labelling service at the victims own edge router which propagates this task throughout the network to all other routers. The triggering of the labelling service is secured by cryptographic authentication of customer and router actions. 
After the configuration of the routers, all messages can be traced back. The traffic data, path labelling data and attack meta data are prepared and stored to cope with the forensic needs of law enforcement. After enough data is collected for the forensic analysis and identification of the attacker, the system can start defence actions. The signature of the malicious packets consists of information like IP addresses, ports or protocol identification as referenced in the IDS alarm is propagated to the routers. They use this signature to start filtering, blocking or delaying such incoming traffic. Furthermore, flow sampling or logging at the routers may be configured according to the attack. However, this is out of scope of the proposed solution. 

The labelling of packets extends the necessary data for digital forensics in a live response scenario. Furthermore, the provider may benefit from these data and the detection vector for blacklisting attackers or bots, identifying botnets within the ISPs network or verify malicious ISPs. In addition, this method provides an approach for detecting Coremelt\footnote{A botnet generate pseudo realistic traffic between individual bots on a specific network node, so that it gets overloaded.} attacks, whereas currently no defence mechanism exists. Another use case for the labelling triggered by the ISP and not the customer may be the verification of load balancing schemes or traffic shaping configured in the network management system.
\section{REQUIREMENTS and ASSUMPTIONS}\label{sec:requirements}
For our analysis we assume that routes can dynamically change during an attack. Furthermore, packets can get lost, and the order of the packets can be changed. Finally, attackers are able to generate any packet and different kinds with faked parameters.
According to the scenario and the various application areas, the following requirements have to be fulfilled. It is mandatory that the system is easy to apply everywhere and that the necessary additional resources are minimal. We address traceback strategies concerning to the following main aspects:
\begin{itemize}
\item \textbf{Single packet traceback} \\ (to detect sophisticated attackers)
\item Detect and differentiate \textbf{multiple attackers}
\item Fast path reconstruction, even during an attack\\ (short \textbf{Attack Detection Time} and fast \textbf{Preventive} actions)
\item Minimal additional network load and performance reduction \\ (\textbf{Efficiency, Costs})
\item \textbf{Traced Hops / Locations} of more than 50 hops\\ (necessary for addressing worst hop connections)
\end{itemize}

For example, a direct call of the website www.torproject.org from the location of the \glqq Universit\"at der Bundeswehr\grqq{} in Munich is transferred over more than 18 routers\footnote{Windows command: tracert www.torproject.org}. This does not include anonymous techniques or other configurations as for example manually chosen proxy servers. The worst case number of nodes traversed by a packet was 39 hops in 1996, the expected worst case hop number in 2011 is about 56 \cite{Chinnery2010}.
\\
Basic aspects of all strategies are \textbf{effectiveness} of the labelling and path reconstruction, \textbf{scalability} and applicability to big environments and \textbf{robustness} of used algorithms. In addition, there should be \textbf{no cooperation necessary} within and between ISPs as well as no complete \textbf{deployment in the entire network}.



\section{RELATED WORK}\label{relatedwork}
Over the past years, many traceback and defence strategies have been developed. So far, a fitting solution for protection or valid forensic identification of attackers has not been developed yet. Table \ref{tab:comparison} evaluates and compares our \textit{Tracemax} system with existing approaches in respect to the identified requirements.\\
Ingress Filtering (IF) \cite{RFC2267} is not a classical IP-traceback algorithm, but is often used as a preventive technique against attacks. This technique prevents IP spoofing and makes it more difficult for an attacker to remain anonymous. It is often used by ISPs and is very important to identify a path as a forensic proof.\\
Router Stamping (RS) consists of two parts \cite{25}. The first part is the marking scheme of the packets; the second is the reconstruction algorithm. The Router Stamping algorithm writes its own information in the Option Field in the IP header. Mainly two marking algorithms exist: Deterministic RS (DRS) and Probabilistic RS (PRS). The DRS is exactly doing what we need for traceback. But the size of the Option Field is very limited. In compliance with the overhead the algorithm is only able to trace up to 9 hops. PRS is an extension of the DRS with a probabilistic component. Every ISP publishes information about their private topology. Multiple attackers as well as DDoS attacks are difficult to differentiate. Due to this, it can take a long time to reconstruct the entire attack tree.\\
The Packet Marking (PM) differentiates between Node-Sampling and Edge-Sampling as marking functions. At the Node-Sampling approach \cite{27} every router writes its own IP address into the forwarding packet and the information is written at the end of the payload part of the IP packet. This can cause failure at the receiver. The PM traceback technique is not efficient with respect to capacity. Furthermore, the marking algorithms can lead to an exceeding of the maximal packet size, which creates further problems and unwanted side-effects. These problems are partly solved by Edge-Sampling \cite{27}. This marking scheme is comparable with PRS and stores the information of a single edge into the packet. But all these functions need a lot of calculation power, which results in large delays. These techniques are difficult to implement and have the same further disadvantages as RS.\\
Another approach is Link Testing (LT). It uses two principles: LT by Input Debugging (LTID) \cite{29} and LT by Controlled Flooding (LTCF) \cite{30}. It try to detect the attack path backwards by generating an overload at specific link. These techniques require a continuous attack for complete attack path identification and prohibit immediate live defence actions. In addition, the router needs a lot of calculation power. For defence actions an automatic communication mechanism between the routers is needed.\\
The ICMP-traceback approach \cite{36} is based on the standardized Internet Control Message Protocol (ICMP). It detects the path of packets backwards by sending packets with a reduced Time-to-Live value. Most routers filter ICMP messages making this technique difficult to apply \cite{34}. The discovered path does not necessarily match the real path of a packet, because of load balancing approaches and other influences.\\
Another approach is the ISP traceback \cite{stelte13}. It writes the ASN number of an ISP in the Option Field of a packet. This allows the traceback to the source ISP of a packet, but not the direct path.\\ There are more algorithms and hybrid solutions to trace packets as discussed in \cite{25}. 

All in all, the known traceback strategies do not fulfill the identified requirements to trace the path of network packets. Most strategies cannot trace enough direct hops. Others need multiple packets or are based on probability.



\section{The new Approach: Tracemax}
Following the scenario from Section  \ref{scenario}, the user has to send a specific request to his ISP to start tracebacking packets. This may be enabled via a service interface or a direct message oriented management access for the customer. This activates the novel traceback algorithm \textit{Tracemax} for a predefined number of routers. These routers have to cooperate and have to be configured for \textit{Tracemax}. Also IF should be activated. This increases the trustworthiness of the IP addresses from the incoming IP packets that will be forwarded by the network.

\textit{Tracemax} consists of a marking scheme and a reconstruction method. These and all other components are explained in details below. The routers are marking the packets on the path during the transmission. The reconstruction method determines the path of a packet afterwards.

\subsection{Marking scheme}
Every router $T_{i}$ on the path writes a predefined $ID_{i-n}$ in the \textit{Option Field} of the IP packet. Even if the size is very limited (40 Bytes), the Option Field is in our case the best choice and the size is sufficient. This increases the network load only slightly and has nearly no side effects. Our defined unique ID number with less than 6 Bits for labeling every port or trunked group of ports of a single active network device is much smaller than 32 Bits in case of an IPv4 address. That way we are able to store many more IDs into the Option Field than with the RS strategy IP addresses. The precise necessary bit size of the ID number depends on the situation and can be adapted. The bit size depends indirectly on the router with most physical connections, which is in the predefined area of \textit{Tracemax}. For performance improvement, single network devices with many ports can be virtually split in multiple devices to lower the maximum necessary bit size for an ID number. The ID can be seen as a lower ISO/OSI layer information, for example the port number of a physical connection on a router. In the process of development, the ID is declared as an abstract number. It is independent from the port number. This allows improvements during the ID assignment and can lead to a reduced necessary bit size for the IDs. Every router writes the assigned and predefined ID of the physical port into the packets Option Field at the outgoing interface. The next router can prove the value and correct it, if necessary.

\subsection{ID assignment}
\begin{figure}[hbt]
\hfill 
     		\begin{minipage} [hbt]{4.192cm}
\begin{center}
\includegraphics[width=1.05\textwidth]{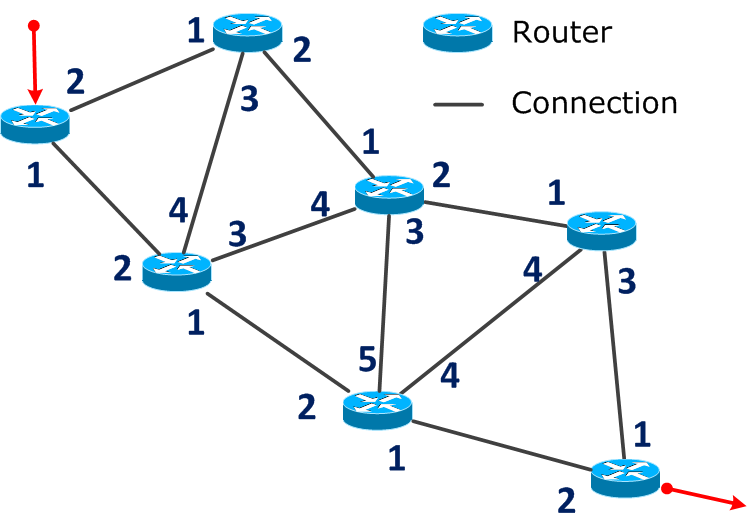}\\
\caption{Valid definition.}
\label{Tracemax2}
\end{center}
 				\end{minipage}
 				\hfill
 				\begin{minipage} [hbt]{4.192cm}
\begin{center}
\includegraphics[width=1.05\textwidth]{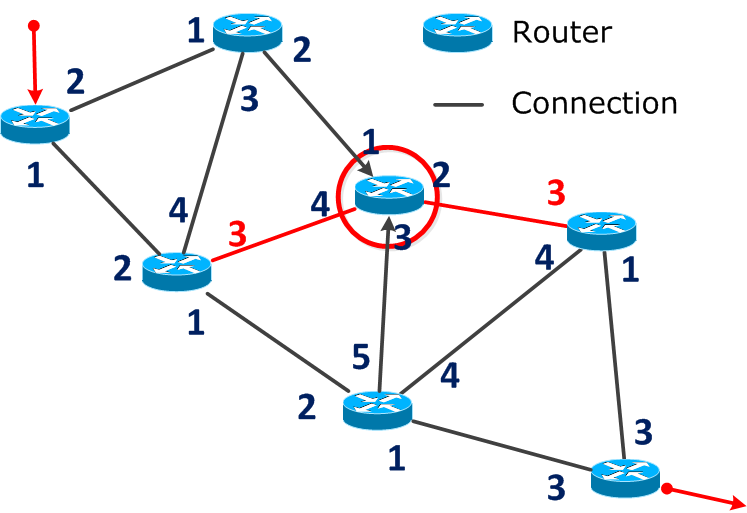}\\
\caption{Invalid definition.}
\label{Tracemaxfail}
\end{center}
 			\end{minipage}
 			\hfill \hfill
 			\end{figure}

Before the \textit{Tracemax} system is rolled out, every physical port of every router gets his own ID, which is not necessarily unique in the entire system. Still the ID numbers are defined in a way that the path reconstruction is uniquely possible. In analogy to the port numbers of a preselected router our algorithm prevents that two neighboring routers get the same ID on direct links to the preselected one. This is achieved by incrementing one of the intended IDs. The ID still requires less bits than an entire IP address. A valid and an invalid example can be seen in Figure \ref{Tracemax2} and \ref{Tracemaxfail}.

In a simple version every router gets his own ID instead of every physical connection. The routers write their router ID in the packet before it gets forwarded. But in this case, often larger numbers are required. The main problem of this simple version is that it is not robust against non-marking routers in between. The traceback information would be only that a packet passed a specific router. But we do not know the incoming and outgoing path. This is the reason to focus on the more complex version with IDs for every physical connection.

The reconstruction function extracts the IDs out of the IP header. The sequence of IDs defines the used connections between the routers and thereby the path of the packet. The function had to know the endpoint where the packet is captured. This information is stored with the packet or the user already knows. In most cases the reconstruction is running at the same system as the endpoint. Otherwise, the closest IP address to the destination could be stored completely into the Option Field as well.

\subsection{Reconstruction scheme}
The reconstruction function matches the sequence of the IDs with the knowledge of the network infrastructure and the predefined IDs. This is done step-by-step backwards, starting from the receiver node. The provider of the network and the administrator of the \textit{Tracemax} have all necessary information allowing the complete reconstruction of the path. The function can also map these routers to their IP addresses. In the end, the order of the routers and the IP addresses of the path are known.

At the borders of different \textit{Tracemax} systems or ISPs the Option Field has to be cleared. This information is not useful for others and it reduces the network load. Such information could be stored at the router or immediately sent to a collector for this path information to analyse the general situation. In the opposite way, it is vice versa. If an incoming packet is entered into the \textit{Tracemax} system, the information in the Option Field has to be cleared. This prevents injecting of false information and generates space for the own traceback information. If an incoming packet at an \textit{Tracemax} system edge router is labeled with the sending routers IP address from the routing table then the path reconstruction is extended with reference to the first external routing device. Since the IDs in the \textit{Tracemax} system are small, the Option Field provides the necessary space.

In summary, the Option Field includes in the complete version the following information:

Preamble : IP-Sender : $1^{st}$ ID : $2^{nd}$ ID : $3^{rd}$ ID : ..... : n. ID : IP-Receiver

A large advantage of \textit{Tracemax} is that the new developed approach does not provide any information about the private topology information of an ISP even if someone gets access to the IDs in the Option Field. The IDs could also be changed in a short interval to avoid reverse engineering. In the scenario of DDoS we have to assume that IP spoofing is used by the attacker. This aspect does not influence our \textit{Tracemax} strategy because it is not based on IP source addresses. Every single packet is marked and can be traced independently from other packets. Hence, variable routes are also detected. In addition, \textit{Tracemax} and the marking information have no influence on the payload and the receiver of such a packet.


\section{Storing the \textit{Tracemax} information at the IP header}\label{Option-Field}
In the following, the Preamble and the entire Option Field is explained in more detail. We mention the construction of the Option Field in the following because not all of the available Bytes in the Option Field are free and flexible usable. The IP header in version 4 is 20 Bytes large and can have an additional Option Field with the variable length (Padding dependent) of up to 40 Bytes.


There exists two different formatting styles to add an information in the \textit{Option Field}. The traceback information can be add as single octet or within a defined option-length. 
\textit{Tracemax} uses the 2nd case to be able to predefine the necessary space.
All option-type parts have to start with: 1 Bit \textit{Copied Flag}, 2 Bits \textit{Option Class}, and 5 Bits \textit{Option Number}.


The 1 Bit \textit{Copied Flag} (CP) defines that this option is copied into all fragments. We do not use this possibility for traceback, so it will be set to $0_{2}$ to keep the network load small.
For the 2 Bit \textit{Class Field} (CF) there exists different valid possibilities. The measurement parameter $10_{2}$ is most fitting for our case.
For the \textit{Option Number}, we have to find a value, which do not create conflicts. \textit{Tracemax} uses the unassigned \textit{Option Number} $1 0110_{2}$ ($22_{10}$).

$1^{st}$ octet: $86_{10}$ = 0x56 = $0 1 0 1 0 1 1 0_{2}$
 
Furthermore, this octet is be followed by \textit{option-length} octet, which specifies the length of the current Option Part including the Option Part header \cite{1981}. The size can be between 0 and 40. In our case, we define the entire option part for our usage.

$2^{nd}$ octet: $40_{10}$ = 0x28 = $0 0 1 0 1 0 0 0_{2}$


It is very important to give attention on this formatting, because of possible malformed packets or the unintentionally mapping with some other option values. For example, the two option values \textit{Loose Source Route (LSR)} and \textit{Strict Source Route (SSR)} are discouraged because of security concerns\footnote{Cisco FAQ. What *is* source routing?, 2014.}. Such packets will be dropped or the option part gets deleted.
This concept can be easily adapted to IP version 6. Therefore, it is necessary to define a new Next Header to store the traceback information.

\section{SIMULATION AND ASSESSMENT}\label{simulation}
The evaluation of the described \textit{Tracemax} system is based on experiments using a prototypic implementation. We have set up a network with multiple virtual machines. Our approach is implemented with the interactive packet manipulation and packet generation tool \textit{Scapy} which is written in Python. The packets are captured with Wireshark. The reconstruction of the path is done manually to analyse possible problems. At the beginning we had issues with malformed packets because of the described formatting of the Option Field in section \ref{Option-Field} and the hex values used by Scapy.
The following Python Script, Listing \ref{lst:python}, describes the marking algorithm of \textit{Tracemax}, which is used at every router. The routers itself are virtual machines during the experiments to get detailed analyses of the system behaviour.
First, the script detects whether the incoming packet has an Option Field or not. Accordingly, the Option Field is added with the Preamble '\textbackslash x56\textbackslash x28'. The Python Script generates a copy of the packet and extends the Option Field with its ID. For simplification, every router has one ID to mark the packets. The ID has exceptional a size of a double hex value because Scapy and to avoid complex Bit transformation for the proof of concept. The \textit{Tracemax} ID of a specific router is defined in variable 'myindex', for example  '\textbackslash x09'. After that, the new generated packet is sent by Scapy to the next network hop. For the simulation, the Script filters on ICMP packets.

\lstset{language=Python} 
\lstset{ %
  backgroundcolor=\color{white},   
  basicstyle=\footnotesize,        
  frame=single,                    
  numbers=left,                    
  numbersep=5pt,                   
  stepnumber=1,                    
}
\begin{lstlisting}[frame=single,caption={Python script for Tracemax.},label=lst:python]
from scapy.all import *

def chgSend(x):
	myindex = '\x09'	# Assigned Port Identifier
	
	optionsarray = x[IP].options 	#Option Field
	if optionsarray.count(1) == 0
	   and str(optionsarray) != '[]':
				optionsstring = str(optionsarray[0])
	else:
				optionsstring = '\x56\x28'

	#Generate packet
	y=IP(src=x[IP].src,dst=x[IP].dst,len=60,options=
			IPOption(optionsstring+myindex))/x[IP].payload
	send(y)

while 1: # Listen to the network interface
	sniff(prn=chgSend, lfilter=lambda x:
	        x.haslayer(ICMP), count=1)
\end{lstlisting}

During the simulation, every virtual machine runs Wireshark. The created captures serves as a reference. The Option Field gets marked and the information from \textit{Tracemax} are usable for the path reconstruction. The experiments validate our concept and show the easy implementation as well as the practicability.

To summarize the results, the following Table \ref{tab:comparison} compares our \textit{Tracemax} system with the existing strategies. This highlights the main advantages of our developed system against existing approaches. The symbols have the following meaning in comparison to the other traceback techniques: + advantage; - disadvantage; o neutral.

With \textit{Tracemax}, we are able to trace the required number of more than 50 hops for just a single packet. The system is still efficient and the attack detection time is short even for multiple attackers. Furthermore, it can be used for a preventive analysis of the network. 

\begin{table}[htb]
	\centering
	\caption{Comparison of different traceback strategies}
	\setlength{\tabcolsep}{2pt}
	\begin{tabular}{|C{1.6cm}|C{1.1cm}|C{0.7cm}|C{0.7cm}|C{0.7cm}|C{0.7cm}|C{0.7cm}|C{0.7cm}|C{0.7cm}|C{0.7cm}|C{0.7cm}|C{0.7cm}|C{0.7cm}|C{0.7cm}|}
		\hline
		\rotatebox{90}{\textbf{Algorithms}} & \rotatebox{90}{\parbox{3cm}{\textbf{Traced Hops\\ or Locations}}} & \rotatebox{90}{\textbf{Effectiveness}} & \rotatebox{90}{\textbf{Efficiency}} & \rotatebox{90}{\textbf{Scalability}} & \rotatebox{90}{\textbf{Costs}} & \rotatebox{90}{\parbox{3cm}{\parbox{3cm}{\textbf{No cooperation\\ necessary}}}} & \rotatebox{90}{\textbf{Robustness}} & \rotatebox{90}{\parbox{3cm}{\textbf{Deploy\\ entire network}}} & \rotatebox{90}{\textbf{Attack detection time}} & \rotatebox{90}{\parbox{3cm}{\textbf{Preventive\\ usable}}} & \rotatebox{90}{\parbox{3cm}{\textbf{Single packet\\ traceback}}} & \rotatebox{90}{\parbox{3.2cm}{\textbf{Multiple\\ attackers}}} \\
		\hline
		RS-DRS & 9 & +  & o & o & + & + & o & + & + & + & + & + \\
		\hline
		RS-PRS & $\infty$ & + & o & o & + & + & o & + & - & + & - & - \\
		\hline
		PM-Node & $\infty$ & + & + & + & o & + & - & o & o & + & + & + \\
		\hline
		PM-Edge & 1 edge & o & o & + & o & + & o & - & - & + & - & - \\
		\hline
		LT-Debug. & $\infty$ & o & - & o & o & - & + & - & - & - & - & - \\ \hline
		LT-Flood. & $\infty$ & - & - & - & + & o & o & + & - & - & - & - \\ \hline
		ICMP     & $<$256 & o & o & + & + & + & - & + & + & - & + & + \\ \hline
		ISP Trace & ISP & - & o & + & + & + & - & + & + & + & + & o \\ \hline\hline
		\textbf{Tracemax} & \textbf{$>$50} & \textbf{+} & \textbf{+} & \textbf{+} & \textbf{+} & \textbf{+} & \textbf{+} & \textbf{o} & \textbf{+} & \textbf{+} & \textbf{+} & \textbf{+} \\ \hline
	\end{tabular}
	\label{tab:comparison}
\end{table}


\section{CONCLUSION}


In this paper we propose \textit{Tracemax}, a novel traceback strategy for tracking and path reconstruction of just single IP packets through the network. It is possible to detect significantly longer paths than with existing methods. The technique does not affect the payload data in the packet, so that labelling information of a traced packet do not need to be cleared before delivery. This makes it very easy to implement and to use. The ISP reveals no private information about the network topology because of abstract IDs. The additional network load is very low, providing good scalability. It is easy to decide between multiple attackers. Variable routes of single packets during a communication can be detected as well.

In further work, concepts for the use of \textit{Tracemax} with competing applications for the \textit{Option Field} are to develop and to investigate the influence of \textit{Multiprotocol Label Switching}. In addition, the need for an entire network deployment should be more flexible, so that it can handle not marking hops. The impact of the packet labelling on the performance of the router has to be analysed in more detail. Finally, the specification of the system is intend to publish in an RFC.

\section*{Acknowledgment}
This work was partly funded by FLAMINGO, a Network of Excellence project (ICT-318488) supported by the European Commission under its Seventh Framework program.

\bibliography{literature}

\end{document}